\documentclass[12pt]{article}
\usepackage[pctex32]{graphics}
\begin{document}
~
~
~
~~
\vspace{2cm}
\begin{center} {\Large \bf  Tubular Solutions of Dirac-Born-Infeld Action on Dp-Brane Background}
                                                  
\vspace{1cm}

                      Wung-Hong Huang\\
                       Department of Physics\\
                       National Cheng Kung University\\
                       Tainan, Taiwan\\

\end{center}
\vspace{1cm}
\begin{center} {\large \bf  Abstract} \end{center}
We use the Dirac-Born-Infeld action on Dp-brane background to find the  tubular  bound state of  a D2 with $m$ D0-branes and $n$ fundamental strings.  The fundamental strings are circular along the cross section of the tube and tube solutions are parallel to the geometry of  Dp-brane background.  Through the detailed analyses we show that only on the D6-brane background could we find the stable tubular solutions.  These tubular configurations may be prevented form collapse by the gravitational field on the curved Dp-brane background.

\vspace{2cm}
\begin{flushleft}
E-mail:  whhwung@mail.ncku.edu.tw\\
\end{flushleft}
%%%%%%%%%%%%%%%%%%%%%%%%%%%%%%%
\newpage
\section {Introduction}
A bunch of strings with D0-branes can be blown-up to a supersymmetric tubular D2-brane which is supported against collapse by the angular momentum generated by the Born-Infeld (BI) electric and magnetic fields [1].  The energy of  the tube is typical of 1/4 supersymmetric configurations, and a calculation confirms that the D2-brane configuration just describes preserves 1/4 supersymmetry, hence the name `supertube'.  More investigations about the supertube solutions have been presented in [2].  The matrix theory interpretation was provided by Bak and Lee in [3] and others [4].  The solutions have also been analyzed from the tachyonic action [5].   

In paper [6]  we have shown that a bunch of  fundamental strings with D0-branes could be bound with D2-brane to form a stable tubular configuration which is prevented from collapsing by the magnetic force in the Melvin background.   The configuration is different from the tube solution of [1] which is supported against collapse by the angular momentum.  

In this paper we will show that the gravitational field on the Dp-brane curved background may support the tubular bound state of $n$ fundamental strings, $m$ D0, with D2, which is denoted as ($n$$F$, $m$D0, D2)-tube, form collapse.  The fundamental strings in here are the circular strings (denoted as $F_c$) along the cross section of tube. (The straight strings along the axial of the tube are denoted as $F_s$.)  When the fundamental strings (and thus the BI electric fields) are chosen to be along the tube cross section (as that in [6]) then the fundamental strings become circular and the tube is stabilized by the gravitational force.  In this case, the circular F-strings are fusing inside the D2 worldsheet by converting itself into homogenous electric flux.  As the direct along the electric is a circle with radius $R$ the open strings now stretch around the circle and the two ends join to each other with a finite probability [7-9].  On the other hand, when the fundamental strings (and thus the BI electric fields) are chosen to be along the axial of the tube (as that in [1]) then the fundamental strings therein is straight and the tube is stabilized by the angular momentum if there are D0-branes (i.e., $m\ne 0$) [1].   The tube solutions considered in this paper are parallel to the space of  Dp-brane background.  We will see that by using the Dirac-Born-Infeld action then, for any values of $(m,n)$, the all possible values of p which allow the stable tubular solutions could be determined. Our results show that only on the D6-brane background could we find the stable tubular solutions.  Note that for the case of p = 6 there are literatures which discussed the existence of other nontrivial configurations [10].

This paper is organized as follow.  In section II we first present the metric of the $N$ coincident Dp-branes [11] and then use the Dirac-Born-Infeld action [12] to calculate the associated Hamiltonian for tubular configurations of ($nF$, $m$D0, D2) bound state.  In section III we present the analyses to find the all possible Dp-branes which allow the stable tubular configurations of  ($nF$, $m$D0, D2) bound state. We make a conclusion  in the last section.

\section {DBI Action on Dp-Brane Background}
The metric, the dilaton $(\phi)$ and the RR field $(C)$ for a system of $N$ coincident $Dp$-branes are given by:
$$ds^2=  H^{-{1\over 2}}_p ~\eta_{\alpha \beta} + H^{1\over 2}_p ~\delta_{ij}, ~~~~~(\alpha, \beta = 0,..,p ; ~ i, j = p+1,..., 9), \eqno{(2.1)}$$
$$e^{2\phi}= H^{{3-p}\over 2}_p, ~~~~~~~ C_{0...p} = \,H^{-1}_p,~~~~~~~H_{p} = 1 + {{N g_s l^{7-p}_s}\over r^{7-p}}, \eqno{(2.2)}$$
where $H_p$ is the harmonic function of $N$ Dp-branes satisfying the
Green function equation in the transverse space [11]. 

 The Dirac-Born-Infeld Lagrangian of tubular bound state of $n$ fundamental strings, $m$ D0, and D2 bound state, for unit tension, is written as [1]
$${\cal S} =  - \int_{V_3} dt\, dz \, d\theta \,e^{-\phi}~\sqrt{- \det (g +F)}\, + \int_{V_3} \, P\left[\sum_s C^{(s)}\right]\,\wedge  e^{2\pi\alpha'F},   \eqno{(2.3)} $$
where $g$ is the induced worldvolume 3-metric, $F$ is the BI 2-form
field strength and $C^{(s)}$ is the s-form RR potential.   $P[C^{(s)}]$ denotes the pullback of the spacetime tensor $C^{(s)}$ to the brane worldvolume.  We take the worldvolume coordinates to be $(t,z,{\theta})$ with ${\theta} \sim {\theta} + 2\pi$.  Then, after fixing the worldvolume for a tubular topology by  the `physical' gauge choice the induced metric of a static straight tube solution of circular cross section with radius $R$ is
$$ ds^2(g) =  H_p^{-1/2} \left(- dt^2 + dz^2\right)+H_p^{1/2} R^2 d{\theta}^2 ,~~~~~ \mbox{if tube is parallel to Dp brane}.~~~~ \eqno{(2.4)} $$
We will allow for a time-independent electric field $E$ and magnetic field $B$ such that the BI 2-form field strength is [1,6]
$$ {\bf F}= E \, dt\wedge dz + B \, dz\wedge d\theta ,~~   ~~~~~ \mbox{if  string  is straight along the axial of the tube}, \eqno{(2.5a)}$$
$$~{\bf F}= E \, dt\wedge d\theta + B \, dz\wedge d\theta,~~~\mbox{if string is circular along the cross section of tube}. \eqno{(2.5b)}$$
Then, from the Dirac-Born-Infeld Lagrangian we can define the momentum conjugate to $E$ 
$$\Pi \equiv {\partial{\cal L}\over \partial E}.   \eqno{(2.6)}$$ 
The corresponding Hamiltonian density is 
$$ {\cal H} \equiv \Pi E - {\cal L}. \eqno{(2.7)}$$
For an appropriate choice of units, the integrals 
$$m \equiv {1\over 2\pi} \oint d\theta \, B , ~~~~~~\mbox{and}~~~~~~n \equiv {1\over 2\pi}\oint d\theta \, \Pi \, , \eqno{(2.8)}$$
are, respectively, the conserved D0-brane charge and IIA string  charge  per unit length carried by the tube [1]. 

   To proceed, we shall discuss the effects of the RR potential on the tube configuration. 

 1.  In our model we have a relation
$$\int_{V_3} \, P\left[\sum_s C^{(s)}\right]\,\wedge  e^{2\pi\alpha'F} = \int_{V_3} \, \left[ P(C^{(3)}) + 2\pi\alpha' \,P (C^{(1)})\,\wedge  F\right].   \eqno{(2.9)}$$
Therefore, although the D2-brane background may provide a 3-form RR potential to give a nonzero contribution to the action, however, as described in figure 1, the coordinate $(t, z, x_1, .., x_{p-1})$, which is that of  the Dp-brane background, and $(t, z, R, \theta)$, which is that of the tube, have only two common coordinate $(t, z)$.  Thus the pullback of the form $C^{(3)}$ is always zero.  This show a fact that the RR potential of the D2-brane background does not affect the tube configuration in our model. 

\vspace{1cm}
\hfil\scalebox{1}{\includegraphics{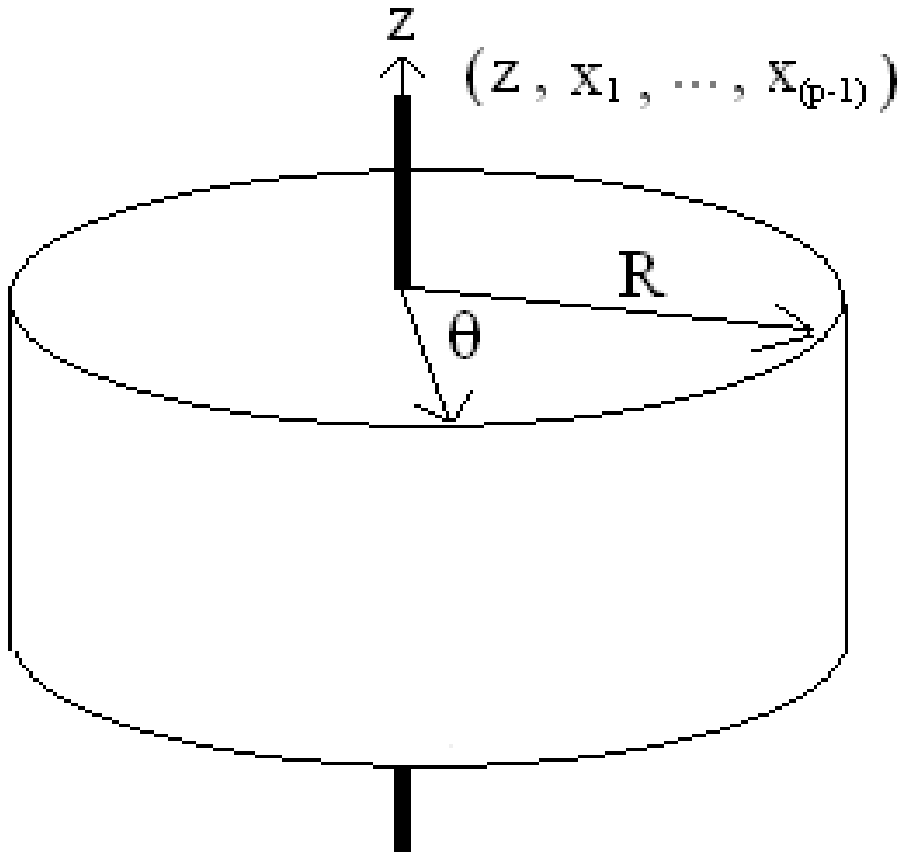}}\hfil\\
{\it ~~~Fig.1. The coordinate $(t, z, x_1, .., x_{p-1})$ is that of  the Dp-brane background and the coordinate $(t, z, R, \theta)$ is that of  the tube.  They have only two common coordinates $(t, z)$.}
\vspace{1cm}

  2.  The Dp-branes are charged electrically and magnetically under RR field strengths.  The respective field strengths are
$$F_{t, x_1,...,x_p, r}^{(p+2)} = \partial_r H^{-1}_p, \eqno{(2.10a)}$$
and 
$$F_{\theta_1, ..., \theta_{(8-p)}}^{(8-p)} = (7-p) (Ng_s l_s)^{7-p}\sqrt{g^{(8-p)}}, \eqno{(2.10b)}$$
in which $g^{(8-p)}$ is the metric of the unit $(8-p)$-sphere transverse to the Dp-brane and coordinates $\theta_1, ..., \theta_{8-p}$ are the spherical coordinates parameterizing the sphere.   Therefore, in the D6-brane background there is a one-form $A_\theta^{(1)}$ which will combine with two-form BI field strength $F_{t,z}$ to contribute the action, i.e. the second term in (2.9).  From (2.5) we see that the BI field strength on ($n$$F_s$, $m$D0, D2)-tubes has $F_{t,z}$ compoment, therefore the tube will feel this RR force under D6-brane background.  On the other hands, the ($n$$F_c$, $m$D0, D2)-tubes does not feel RR force.  In this paper, as we want to see whether the pure gravitational field could stabilize a tube  configuration we will therefore in below consider only the ($n$$F_c$, $m$D0, D2)-tubes on the Dp-Brane Background.

  3. As the tube we considered will along one axial  (i.e. $z$) of the Dp-brane, as described in figure 1,  the possible values of $p$ are 2, 4, and 6.  These are the Dp-brane backgrounds considered below. 

   Then, using the above formula and after the calculations we have the following Langrangian for a ($nF_c$, $m$D0, D2)-tube:
\\
\\
$~~~~~ L = - \, H_p^{p-3\over 4}\, \sqrt {-Det \left(\begin{array}{ccc}
-H_p^{-1/2}         &  0             & E      \\
0             & H_p^{-1/2}           & B    \\

-E             & -B             & H_p^{1/2} R^2   \\
\end{array}\right)}\, = - \, H_p^{p-4\over 4}\,\sqrt {R^2 + B^2 - E^2},$
\\
\\
$${\cal H} = \,\sqrt {\left(R^2 + B^2\right) \left( H_p^{p-4\over 2} + \Pi^2\right)}. \hspace{8cm}\eqno{(2.11)} $$
\\
In the following sections we will use the above formulas to find the all possible values of $p$ which allow the stable configurations of ($nF_c$, $m$D0, D2)-tube.    
\\
\section{Stable Tubular Configurations}
We present in this section the all tube solutions.
\subsection{(D2)-tube}
The first stable tube we are searching is constructed simply by D2.   In this case the tube is denoted by (D2)-tube.   Using (2.11) we have the relations:
$${\cal H}_{(D2)-tube} =  \left\{\begin{array}{ccc}
\sqrt{R^7\over R^5 + N g_s l^{5}_s} \, ,             & p = 2&       \\
R,                  & p = 4    & \\
\sqrt{R^2 +  R N g_s l_s}  \, ,                   & p = 6.   & \\
\end{array} \right . \eqno{(3.1)} $$
These energy densities are increasing functions of $R$ and there does not have a stable tube with finite radius.   Thus we conclude that there is no stable (D2)-tube  on any Dp-Brane background. 

\subsection{($n F_c$, D2)-tube}
The next stable tube we are searching is constructed by D2 and circular string $F_c$.   In this case the tube is denoted by ($n F_c$, D2)-tube.   Using (2.11) we have the relations:
\\
$${\cal H}_{(n  F_c , D2)-tube} =  \left\{\begin{array}{ccc}
\sqrt{R^2 \left({R^5\over R^5 + N g_s l_s^5} + \Pi^2\right)} \, ,  & p = 2 ,&  \\
\sqrt{R^2 \left(1+ \Pi^2\right)} \, ,           & p = 4 ,   & \\
\sqrt{R^2 \left({R + N g_s l_s\over R} + \Pi^2\right)} \, , & p = 6.   & \\ \end{array} \right. \eqno{(3.2)} $$
\\
The above energy densities are increasing functions of $R$ and there does not have a stable tube with finite radius.   Thus we conclude that there is no stable  ($F_c$, D2)-tube on any Dp-Brane background.

\subsection{($m$D0, D2)-tube}
In this subsection we search the tube constructed by $m$D0 with D2.   In this case the tube is denoted by ($m$D0, D2)-tube.   Using (2.11) we have the relations:
$$~~{\cal H}_{(mD0 , D2)-tube} = H_p^{p-4\over 4} \sqrt{R^2 + B^2}= \left\{\begin{array}{ccc}
\sqrt{(R^2 + B^2)\,\left(R^5\over R^5 + N g_s l^{5}_s\right)} \, ,      & p = 2 ,&       \\
\sqrt{\left(R^2 + B^2\right)},           & p = 4 ,    & \\
\sqrt{\left(R^2 + B^2\right)\, \left(1 + { N g_s l_s\over R}\right)} \, ,                   & p = 6 .   & \\
\end{array} \right . ~~~~~\eqno{(3.3)} $$
The energy densities in the cases of p=2 and p=4 are increasing functions of $R$ and thus there is no stable ($m$D0, D2)-tube on D2- or D4-brane background.

   On the other hand, in the case of  p = 6 the energy density ${\cal H}\rightarrow \,   \infty $  as $R \rightarrow\, 0$ .   This implies that the tube is stable with a finite radius.  Thus we conclude that the $(mD0 , D2)$-tube can be stabilized by the gravitational field from the background of D6 when the tube is parallel to the space of  D6 background.   In figure 2 we plot the radius dependence of  the energy density.   We see in there that there is a finite radius $R_*$ at which the energy density becomes a minimum.  

\vspace{1cm}
\hfil\scalebox{1}{\includegraphics{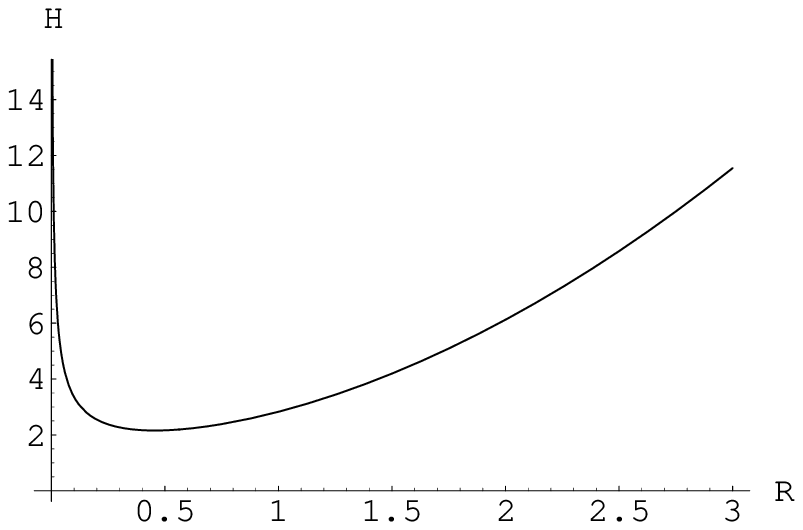}}\hfil\\
{\it ~~~Fig.2. Radius dependence of  the energy density ${\cal H}(R)$ in (3.4) for the case of p = 6. The units are taken to be $N g_s l_s = 1$.  There is a finite radius at which the energy density becomes a minimum.}
\vspace{1cm}
\\
In figure 3 we also plot the $N$-dependent of the tube radius $R_*$.   We see that if $N=0$ then there is no stable tube with finite radius.  However, if we  turn on the gravitational field of background D6-Brane then the gravitational force in the space could support the $(mD0 , D2)$-tube from collapse into zero radius.   We also see that the tube radius is an increasing function of  $N$, the number of the D6-brane on the background geometry.  This means that the background with larger value of $N$, which represents that there is the larger number of the D6-brane on the background geometry, will give a larger force to support the tube from collapse into zero radius.   Thus the radius of the tube will be getting larger.   This is an interesting property found in this paper.

\vspace{1cm}
\hfil\scalebox{1}{\includegraphics{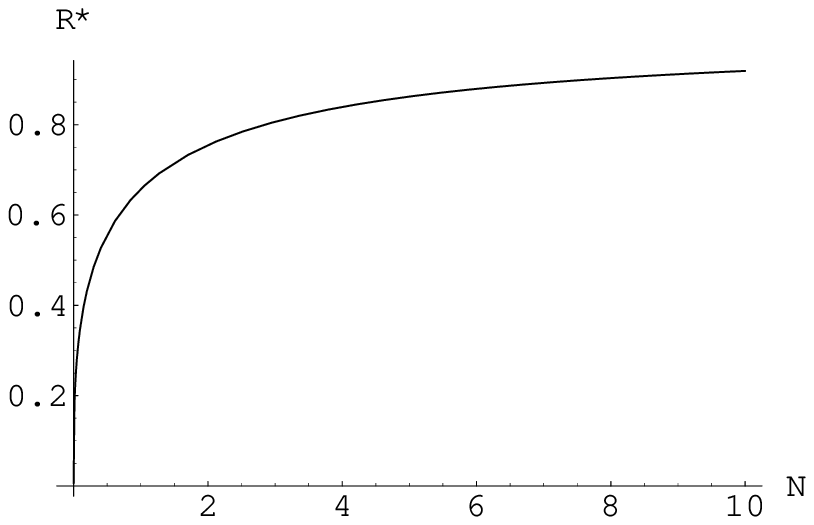}}\hfil\\
{\it ~~~Fig.3. $N$ dependence of  tube radius $R_*$.  The units are taken to be $g_s l_s = 1$.  The figure shows that the tube radius is an increasing function of  $N$.}
\vspace{1cm}
\subsection{($n$$F_c$, $m$D0, D2)-tube}
In this subsection we search the tube of the bound state of $n$$F_c$, $m$D0 with D2.   In this case the tube is denoted by ($n$$F_c$, $m$D0, D2)-tube.  Using (2.11) we have the relations:
$$~~~{\cal H}_{(nF_c, mD0, D2)-tube} = \left\{\begin{array}{ccc}
\sqrt{\left(R^2  + B^2\right)\left({R^5\over R^5 + N g_s l_s^5} + \Pi^2\right)} \, ,            & p = 2 ,&  \\
\sqrt{\left(R^2  + B^2\right)\left(1+ \Pi^2\right)} \, ,           & p = 4 ,   & \\
\sqrt{\left(R^2  + B^2\right)\left({R + N g_s l_s\over R} + \Pi^2\right)} \, , & p = 6.   & \\ \end{array} \right. \eqno{(3.4)} $$
As same as the discussions of (3.3) the above equation implies that only  on D6-brane background could we find a stable ($n$$F_c$, $m$D0, D2)-tube with finite radius.

The above solutions are the all possible values of p which allow the stable tubular solutions.  As the energy densities of these tubular configurations become infinity as $R\rightarrow 0$, they have a finite radius at which the configurations are stable.   This means that the gravitational field of the curved Dp-brane background may prevent these tubular configurations form collapse.

\section{Conclusion}
The supertubes first found in [1] are the bunch of strings with D0-branes which are blown-up to a tubular D2-brane.   They are supported against collapse by the angular momentum.   In paper [6] we have shown that a bunch of circular fundamental strings with D0-branes could be bound with D2-brane to form a stable tubular configuration which is prevented from collapsing by the magnetic force in the Melvin background. 

In this paper we have shown that the gravitational field on the Dp-brane curved background may support some tubular configurations of $n$ fundamental circular strings $F_c$, $m$ D0, and D2 bound state form collapse.  Through the detailed analyses we have found that the possible stable tubes are the ($m$D0, D2)-tube, ($n$$F_c$, $m$D0, D2)-tube on the D6-brane background.  

It hopes that these investigations may help us to understand the effects of the nontrivial background on the nontrivial brane configurations.   Finally, it is interseting to investigate the properties of the tubular configurations on the Dp-brane background by using the non-abelian Dirac-Born-Infeld action [10] or matrix theory [3,4].  These tubes may also be relevant to the brane world and string cosmology.  These problems remain to be studied. 

\newpage
 
%%%%%%%%%%%%%%%%%%%%%%%
\begin{center} {\large \bf  References} \end{center}
\begin{enumerate}
\item D. Mateos and P. K. Townsend, ``Supertubes'', Phys. Rev. Lett. 87 (2001) 011602 [hep-th/0103030];\\
 R. Emparan, D. Mateos and P. K. Townsend, ``Supergravity Supertubes'', JHEP 0107 (2001) 011 [hep-th/0106012];\\
 D.~Mateos, S.~Ng and P.~K.~Townsend, ``Tachyons, supertubes and brane/anti-brane systems'', JHEP  0203 (2002) 016 [hep-th/0112054];\\
 P. K. Townsend, ``Surprises with Angular Momentum'', Annales Henri Poincare 4 (2003) S183 [hep-th/0211008].
\item  M. Kruczenski, R. C. Myers, A. W. Peet, and D. J. Winters,``Aspects of supertubes'',  JHEP 0205 (2002) 017 [hep-th/0204103];\\
Y. Hyakutake and N. Ohta, ``Supertubes and Supercurves from M-Ribbons,'' Phys. Lett. B539  (2002) 153 [hep-th/0204161]; \\
N. E. Grandi and A. R. Lugo, ``Supertubes and Special Holonomy'', Phys. Lett. B553 (2003) 293 [hep-th/0212159];\\
 B. Cabrera Palmer and D. Marolf , `` Counting Supertubes'',  JHEP 0406 (2004) 028 [hep-th/0403025]; \\
D. Bak, Y. Hyakutake, and N. Ohta, ``Phase Moduli Space of Supertubes,'' [hep-th/0404104];\\
Wung-Hong Huang, ``Condensation of Tubular D2-branes in Magnetic Field Background'' Phys. Rev. D70 (2004) 107901[hep-th/0405192].
\item D. Bak, K. M. Lee, ``Noncommutative Supersymmetric Tubes'',  Phys. Lett. B509 (2001) 168 [hep-th/0103148].
\item D. Bak and S. W. Kim, ``Junction of Supersymmetric Tubes,'' Nucl. Phys.  B622 (2002) 95 [hep-th/0108207];\\
 D. Bak and A. Karch, ``Supersymmetric Brane-Antibrane Configurations,'' Nucl. Phys. B626 (2002) 165 [hep-th/011039];\\
 D. Bak and N. Ohta, ``Supersymmetric D2-anti-D2 String,'' Phys. Lett.  B527 (2002) 131 [hep-th/0112034];\\
 D. Bak, N. Ohta and M. M. Sheikh-Jabbari, ``Supersymmetric Brane-Antibrane Systems: Matrix Model Description, Stability and Decoupling Limits,'' JHEP  0209 (2002) 048 [hep-th/0205265].
\item C. Kim, Y. Kim, O-K. Kwon, and P. Yi, ``Tachyon Tube and Supertube,''  JHEP 0309 (2003) 042 [hep-th/0307184]; \\
Wung-Hong Huang, ``Tachyon Tube on non-BPS D-branes,'' JHEP 0408 (2004) 060 [hep-th/0407081].
\item Wung-Hong Huang, ``Tube of (Circular F, D0, D2) Bound States in Melvin  Background'' Phys. Lett. B599 (2004) 301 [hep-th/0407230].
\item E. Witten, ``Bound States of Strings and p-Branes,'' Nucl.\ Phys.\ B460 (1996) 335 [hep-th/9510135].
\item  C.G.~Callan, I.R.~Klebanov,  ``D-brane Boundary State Dynamics,'' Nucl. Phys.  B465 (1996) 473-486  [hep-th/9511173].
\item  N.~Seiberg, L.~Susskind and N.~Toumbas, ``Strings in background electric field, space/time non-commutativity and a new noncritical string theory,''
JHEP 0006 (2000) 021 (2000) [hep-th/0005040]; \\
I.R.~Klebanov and  J.~Maldacena, ``1+1 Dimensional NCOS and its U(N) Gauge Theory Dual'', I. J. Mod. Phys. A16 (2001) 922  [hep-th/0006085]; \\
U. H. Danielsson, A. Guijosa, ans M. Kruczenski, ``IIA/B, Wound and Wrapped,'' JHEP 0010 (2000) 020 [hep-th/0009182].
\item Y. Hyakutake, ``Gravitational Dielectric Effect and Myers Effect,''  [hep-th/0401026]; Wung-Hong Huang, ``Fuzzy Rings in D6-Branes and Magnetic Field Background,'' JHEP 0407 (2004) 012 [hep-th/0404202]; K.  L. Panigrahi, ``D-Brane Dynamics in Dp-Brane Background,'' Phys.Lett. B601 (2004) 64 [hep-th/0407134].
\item G.~W.~Gibbons and K.~Maeda, ``Black holes and membranes in higher dimensional theories with dilaton fields,'' Nucl.\ Phys.\ B298 (1988) 741.
\item  R. G. Leigh, ``Dirac-Born-Infeld Action From Dirichlet Sigma Model'', 
Mod. Phys. Lett.  A4 (1989) 2767.

\end{enumerate}
\end{document}